\documentclass{appolb}
\usepackage{epsfig}
\usepackage{rotating}
\usepackage{amsmath}
\usepackage{psfrag}
\allowdisplaybreaks[1]

\usepackage{subfigure}

\def\theequation{\thesection.\arabic{equation}}

\newcommand{\Gev}{\, {\rm GeV}}

\newcommand{\bea}{\begin{eqnarray}}
\newcommand{\eea}{\end{eqnarray}}
\newcommand{\bd}{\begin{displaymath}}
\newcommand{\ed}{\end{displaymath}}

\newcommand{\be}{\begin{equation}}
\newcommand{\ee}{\end{equation}}
\newcommand{\bi}{\begin{itemize}}
\newcommand{\ei}{\end{itemize}}
\newcommand{\ord}{{\cal O}}

\headtitle{PHOTON STRUCTURE FUNCTIONS: 1978 AND 2005}
\headauthor{Andrzej J. Buras}
\begin{document}

%%%%%%%%%%%%%%%%%%%%%%%%%%%%%%%%%

\thispagestyle{empty}
\phantom{xxx}
\vskip1truecm
\begin{flushright}
 TUM-HEP-616/05 \\
December 2005
\end{flushright}
\vskip1.0truecm
\centerline{\LARGE\bf PHOTON STRUCTURE FUNCTIONS: }
\vskip0.3truecm
\centerline{\LARGE\bf  1978 AND 2005}
   \vskip1truecm
\centerline{\Large\bf Andrzej J. Buras}
\bigskip
\centerline{\sl Technische Universit{\"a}t M{\"u}nchen}
\centerline{\sl Physik Department} 
\centerline{\sl D-85748 Garching, Germany}
\vskip1truecm
\centerline{\bf Abstract}
I describe the early days of the photon structure functions.
In particular I discuss the parton model result of Walsh and 
Zerwas (1973), leading order QCD calculation of Witten (1976) and 
next-to-leading QCD
calculation of Bardeen and myself (1978). A very brief summary
of the progress made from 1978 until 2005 is also given.

\vskip1truecm

\centerline{\it  Invited Talk  given at }
\centerline{\bf ``The Photon: its First Hundred Years and the Future"}
\centerline{\it Warsaw, August 30 -- September 03, 2005}
%%% end title page %%%%%%%%%%%%%

\newpage
\mbox{}
\thispagestyle{empty}
\newpage
\setcounter{page}{1}

%%%%%%%%%%%%%%%%%%%%%%%%%%%%%%%%%

\eqsec
\title{PHOTON STRUCTURE FUNCTIONS: 1978 AND 2005}
\author{Andrzej J. Buras
\address{Technische Universit{\"a}t M{\"u}nchen,
  Physik Department\\ D-85748 Garching, Germany}}
\maketitle

\begin{abstract}
I describe the early days of the photon structure functions.
In particular I discuss the parton model result of Walsh and 
Zerwas (1973), leading order QCD calculation of Witten (1976) and 
next-to-leading QCD
calculation of Bardeen and myself (1978). A very brief summary
of the progress made from 1978 until 2005 is also given.
\end{abstract}

%%% MAIN TEXT

\section{Introduction}
I have been asked to review the early days of the photon
structure functions and to summarize the present status of
this field. I have worked actively on deep inelastic proton
and photon structure functions \cite{Buras:1979yt} 
from 1977 to 1981, but, after
summarizing the status of perturbative QCD at the 
Photon-Lepton Symposium in Bonn in 1981, I moved to study
technicolour models, flavour physics, weak, CP-violating and rare decays of 
K-and B-mesons, supersymmetry, extra dimensions, little Higgs and
petite unification. Returning in 2005 to photon structure
functions, after being decoupled from this field for 24
years, was a very interesting experience. One should realize
that in 1981 no experimental data on photon structure 
function $F_2^\gamma$ were available,
although the theory had reached already a rather advanced stage.
I felt like a person returning from a long journey to the
earth. My main question was whether predictions for $F_2^\gamma$, in
which I took part mainly in the second half of 1978, have been 
confirmed by the
future data. I will give the answer to this question at the
end of this writing.

 \section{Simple Parton Model Result}
\begin{figure}[hbt]
\vspace{0.10in}
\centerline{
\epsfysize=1.7in
\epsffile{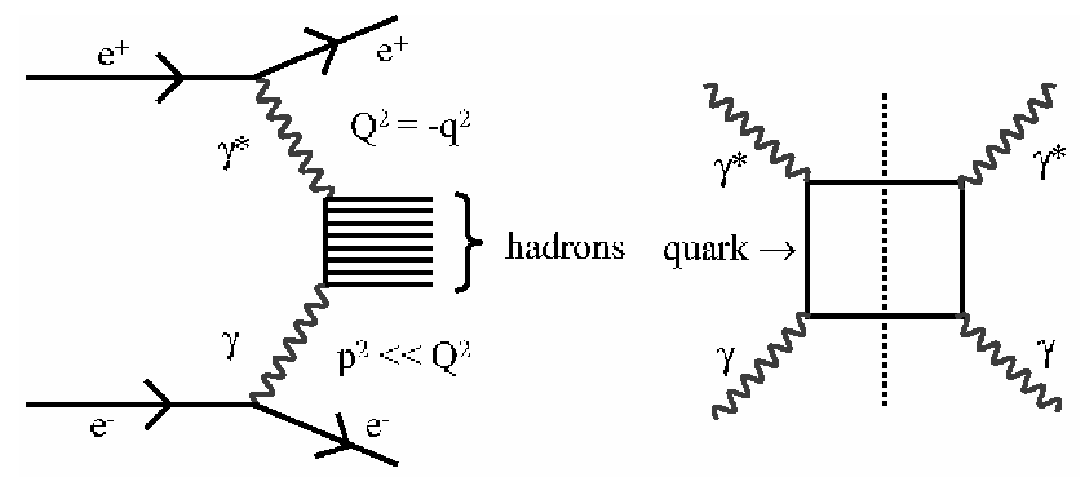}
}
\vspace{0.08in}
\caption{The basic process and the simple Parton Model.}\label{FIG1}
\end{figure}

The story begins in the early 1970's, when Brodsky,
Kinoshita, Terezawa (1971), Walsh (1971), Walsh and Zerwas
(February 1973) and Kingsley (May 1973) analyzed the deep
inelastic scattering of a highly virtual photon $(Q^2\gg p^2)$ on a
real $(p^2\approx 0)$ photon target as seen in Fig.~\ref{FIG1}. 
If the photon behaved only as a vector
meson, the corresponding structure function $F_2^\gamma(x,Q^2)$ would exhibit
$Q^2$ and Bjorken $x$ dependences similar to the one of the proton structure
function $F_2^p(x,Q^2)$. But as pointed out by Walsh and Zerwas 
\cite{Walsh:1973mz}, at very
large $Q^2$ the point-like contribution to $F_2^\gamma(x,Q^2)$, represented by
the box diagram in Fig.~\ref{FIG1}, should dominate over the hadronic
component. Evaluating this box diagram they found
\begin{equation}\label{P1}
F_2^\gamma(x,Q^2)=\tilde F_2^\gamma(x) \log\frac{Q^2}{\Lambda^2}+ ...,
\end{equation}
where
\begin{equation}\label{P2}
\tilde F_2^\gamma(x)=\frac{\sum e^4_i}{16 \pi^2}~x(1-2x+2x^2)
\end{equation}
with $e_i$ being quark charges. I have introduced a scale $\Lambda$ for 
convenience and the non-logarithmic terms in (\ref{P1})
can be found in the original paper.

The most important results in (\ref{P1}) and (\ref{P2})
are:
\begin{itemize}
\item
 $F_2^\gamma(x,Q^2)$ {\it increases} at fixed $x$ logarithmically with $Q^2$ as
       opposed to the corresponding {\it decrease} (at not too small $x$)
       observed for $F_2^p(x,Q^2)$.
\item
  The $x$-dependence of  $F_2^\gamma(x,Q^2)$ at large $Q^2$ is fully calculable
as opposed
to  $F_2^p(x,Q^2)$, where only the $Q^2$ dependence (scaling violations) can
be predicted within renormalization group improved perturbation theory
but the actual shape of  $F_2^p(x,Q^2)$ at a given $Q^2$ can only be found
by means of non-perturbative methods or directly from the data.
\end{itemize}
\boldmath
\section{Master Formula for $F_2^\gamma(x,Q^2)$ in QCD}
\unboldmath
The simple parton model result of Walsh and Zerwas has been
generalized by Witten (1976) \cite{Witten:1977ju} to QCD in the leading
logarithmic approximation (LO). The NLO corrections to
Witten's result have been calculated two years later by
Bardeen and myself (1978) \cite{Bardeen:1978hg} but the complete NNLO
contributions have been presented for the first time only at
this conference \cite{VOGT}, that is 27 years later.

The master formula for the moments of $F_2^\gamma(x,Q^2)$ in QCD reads as
follows \cite{Bardeen:1978hg}:
\begin{eqnarray}
\lefteqn{\int^1_0 dx x^{n-2} F_2^\gamma(x,Q^2)=}
\nonumber\\
&&\alpha^2\left[\frac{4\pi}{\beta_0\alpha_s(Q^2)} a_n + b_n 
+\sum_{i=1}r_n^{(i)}(\alpha_s(Q^2))^i
+\sum_{i=1}h_n^{(i)}(\alpha_s(Q^2))^{d_n^{(i)}}\right],
\label{P3}\,.
\end{eqnarray}
where n=2,3,.., $\alpha$  is the QED structure constant, $\alpha_s(Q^2)$ is the
QCD structure constant and $\beta_0$ the famous LO $\beta$-function
coefficient for which Gross, Politzer and Wilczek received
the Nobel-Prize last year \cite{ASF}. It should be stressed that
\begin{itemize}
\item
 $a_n$, $b_n$ and $r_n^{(i)}$ can be calculated in perturbation theory and
       correspond to LO, NLO and NNLO for $i=1$, respectively.
\item
The coefficients $h_n^{(i)}$ in the last sum are uncalculable in
perturbative QCD. They can be calculated in the vector
dominance model or in principle by means of
lattice methods. They can also be extracted from the data.
\item 
As $d_n^{(i)}\ge 0$, for very large $Q^2$ the first two terms dominate,
except for one of $n=2$ for which $d_2^{(i)}=0$ and 
the $Q^2$ independent piece from
the last sum 
in (\ref{P3}) has to be added
to $b_2$.
\end{itemize}

 \section{More on Witten's Analysis}
Let us recall the basic formula for the moments of the
proton structure function:
\begin{equation}\label{P4}
\int^1_0 dx x^{n-2} F_2^p(x,Q^2)=
\sum_{i=NS,\psi,G}
C^i_n(\frac{Q^2}{\mu^2},g^2)\langle p| O^n_i| p\rangle.
\end{equation}

Here $C^i_n$ denote the Wilson coefficients (calculable in RG
improved perturbation theory) of the local operators $O_{NS}^n$, $O_{\psi}^n$ and 
$O_{G}^n$ (quark non-singlet, quark singlet, gluon) and 
$\langle p| O^n_i| p\rangle$ the corresponding 
non-perturbative matrix elements between the proton states.

As pointed out by Witten \cite{Witten:1977ju}, in the case of $F_2^\gamma$,
an additional
operator, $O_{\gamma}^n$, the analog of the gluon operator $O_{G}^n$ with the
non-Abelian field strength tensor $G_{\alpha\beta}$ replaced by the
electromagnetic tensor $F_{\alpha\beta}$, has to be taken into account. The
reason is that, although the Wilson coefficients $C_n^\gamma$ are 
$\ord(\alpha)$, the matrix elements $\langle \gamma| O^n_\gamma| \gamma\rangle$
are $\ord(1)$. Therefore, the $O^n_\gamma$ 
contribution to $F_2^\gamma$ is of the same order in $\alpha$ as the
contribution of quark and gluon operators. The latter have
Wilson coefficients $\ord(1)$, but matrix elements in photon
states $\ord(\alpha)$.

Explicitly:
\begin{equation}\label{P5}
\int^1_0 dx x^{n-2} F_2^\gamma(x,Q^2)=
\sum_{i=NS,\psi,G}
C^i_n(\frac{Q^2}{\mu^2},g^2)\langle \gamma| O^n_i| \gamma\rangle+
C^\gamma_n(\frac{Q^2}{\mu^2},g^2,\alpha)
\langle \gamma| O^n_\gamma| \gamma\rangle
\end{equation}
with $\langle \gamma| O^n_\gamma| \gamma\rangle=1$.

Including only LO contributions Witten found
\begin{equation}\label{P6}
\int^1_0 dx x^{n-2} F_2^\gamma(x,Q^2)=
C^\gamma_n(\frac{Q^2}{\mu^2},g^2,\alpha)
=\alpha^2\frac{4\pi}{\beta_0 \alpha_s(Q^2)} a_n=\alpha^2 a_n
\log\frac{Q^2}{\Lambda^2}.
\end{equation}
Thus for very large $Q^2$ the logarithmic behaviour in $Q^2$ found
by Walsh and Zerwas in the simple parton model remains valid
in QCD as expected.  
However, what Witten found is that the coefficients
$a_n$ of the leading logarithm differ from the moments of 
$\tilde F_2^\gamma(x)$ in (\ref{P2}).
The anomalous behaviour of $F_2^\gamma(x,Q^2)$ in
QCD can be traced back to the non-vanishing anomalous
dimension matrix in the space of the operators 
$O_{NS}^n$, $O_{\psi}^n$ and 
$O_{G}^n$,
known already from proton deep inelastic scattering and
from the anomalous dimension $K^n_G$, describing the mixing of 
$O_{\gamma}^n$ and 
$O_{G}^n$, under renormalization. Setting all these anomalous
dimensions to zero and keeping only the mixing of 
$O_{NS}^n$ and 
$O_{\psi}^n$
with $O_{\gamma}^n$, Witten's result reduces to the one of Walsh and
Zerwas: the {\it anomalous behaviour} of photon structure functions 
originates in certain non-vanishing {\it anomalous} dimensions.
Witten's result has been rederived by Llewellyn Smith (1978) using 
a diagrammatic method and by DeWitt et al.(1979) and Brodsky et al. (1978)
 in 
the framework of the Altarelli-Parisi approach (DGLAP).

\begin{figure}
%\vspace*{0.3truecm}
\begin{center}
\includegraphics[width=6cm]{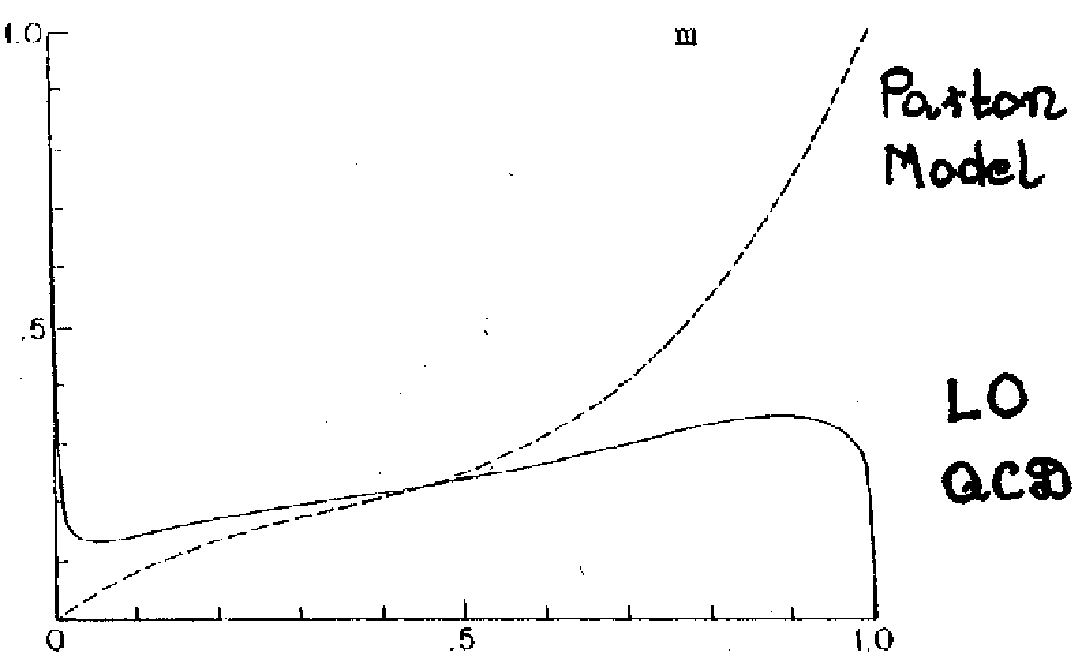}
\includegraphics[width=5cm]{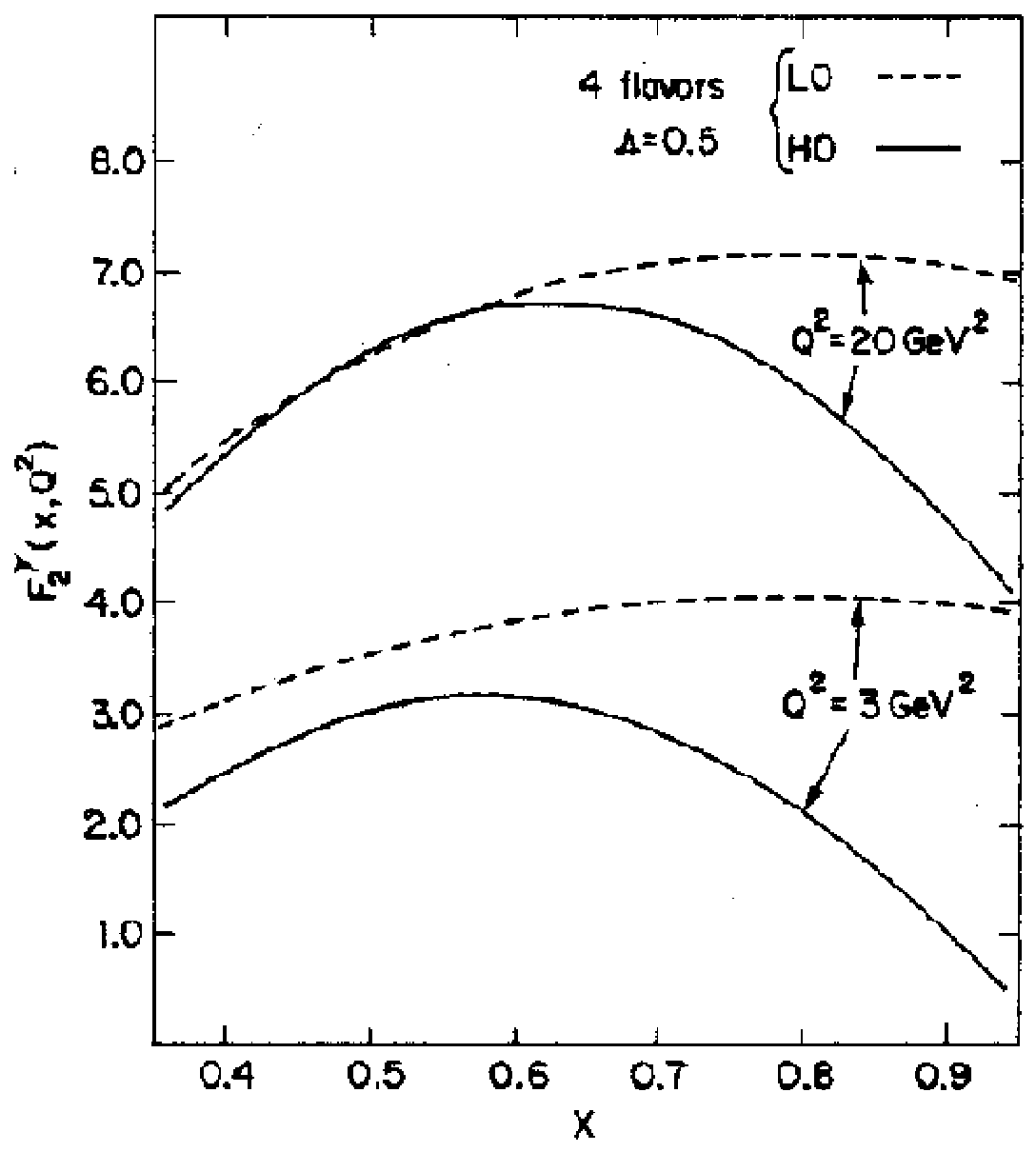}
\end{center}
\caption{$F_2^\gamma$ in the Parton Model \cite{Walsh:1973mz}, QCD at LO 
\cite{Witten:1977ju} and NLO \cite{Bardeen:1978hg}.}\label{FIG2}
\end{figure}

In summary the most important two results of Witten's paper
are the following ones:
\begin{itemize}
\item
 The coefficient of $\log Q^2$ is calculable in QCD. That is
 $F_2^\gamma(x,Q^2)$ is calculable in QCD at large $Q^2$.
\item
It differs from the free parton model result by finite
calculable factors. In particular for $x$ approaching 1  the
increase of $F_2^\gamma(x,Q^2)$ with $x$, observed also in QCD at moderate $x$,
turns into a decrease as shown in Fig.~\ref{FIG2}.
\end{itemize}

 \section{Our 1978 NLO Analysis}
My own work in this field has been triggered by the visit of
Jonathan Rosner to Fermilab in the summer of 1978.
Jonathan wanted to know more about Witten's paper and I
explained it to him in the manner presented above. But on
24th of September 1978 I realized that Witten's result could
be generalized to NLO without doing a single Feynman diagram
calculation. One had only to translate the elements of the
NLO analysis for $F_2^p(x,Q^2)$ into those for $F_2^\gamma(x,Q^2)$. 
A complete NLO
analysis of $F_2^p(x,Q^2)$ has been presented for the first time in June
1978 in a work in collaboration with Bardeen, Duke and Muta 
\cite{Bardeen:1978yd}.
It included the earlier result for two-loop anomalous
dimension matrix involving 
$O_{NS}^n$, $O_{\psi}^n$ and 
$O_{G}^n$  of Floratos, Ross
and Sachrajda (1977,1978). Thus in the fall of 1978 it was indeed
possible without too much work to generalize Witten's result
to NLO. I was truly delighted, when Bill Bardeen agreed to
join me in our second project \cite{Bardeen:1978hg}.

The main virtues of our work, that provided the constants $b_n$
in (\ref{P3}) in QCD were the following ones:
\begin{itemize}
\item
   Having $b_n$ at hand, a meaningful determination of 
   $\alpha_s^{\overline{MS}}$ or $\Lambda_{\overline{MS}}$ 
    \cite{Bardeen:1978yd}
      from $F_2^\gamma(x,Q^2)$ became possible. In view of the calculability of
       $F_2^\gamma(x,Q^2)$ at large $Q^2$, the prospects for the 
      determination of $\alpha_s^{\overline{MS}}$ from
       $F_2^\gamma(x,Q^2)$ appeared at least in principle better than 
   from $F_2^p(x,Q^2)$. We will return to it at the end of this writing.
\item  
Improved accuracy for the prediction of the shape of $F_2^\gamma(x,Q^2)$ 
in QCD.
\item
Identification of large NLO corrections for large $n$ or
equivalently large $x$, that made the turn over in the $x$-dependence
of $F_2^\gamma(x,Q^2)$ stronger than at LO as seen in Fig~\ref{FIG2}.
\item
First comments on the divergent behaviour of the 
point-like component for $x\to 0$ that has been analyzed by many other
authors in the 1980's.
\end{itemize}

In the following years our result has been confirmed by many
groups, in particular by Duke and Owens \cite{Duke:1980ij}, 
Gl\"uck, Grassie and Reya \cite{GR81} and
others, but in 1991 Fontannaz and Pilon \cite{Fontannaz:1992gj}
 and indepently
Gl\"uck, Reya and Vogt \cite{Gluck:1991ee} spotted a small error in our
translation from the mixing $(O_G^n,O^n_G)$ relevant for 
$F_2^p(x,Q^2)$
to the mixing 
$(O_G^n,O^n_\gamma)$ relevant for $F_2^\gamma(x,Q^2)$.
This was a
very stupid error but fortunately without essentially any
numerical consequence for $F_2^\gamma(x,Q^2)$. A visible impact on the gluon
distribution in the photon has been however identified.

 \section{Photon Stucture Functions $1978-2005$}
There is certainly no space to describe here in an adequate
manner the developments after 1978. Selected reviews can be
found in \cite{Nisius,Krawczyk:2000mf,Godbole:2003ym}. Let me then list only 
a few points:

On the theoretical side:
\begin{itemize}
\item
The evolution equations for quark and gluon
       distributions in the photon have been studied by Brodsky et
       al (1978), Gl\"uck, Grassie and Reya (1983), Rossi (1983), Drees (1983)
       and others. See additional references below.
\item
The issue of the singular behaviour for $x\to 0$ has been
addressed in several papers by Duke and Owens, Bardeen,
Gl\"uck, Grassie and Reya, Rossi, Antoniadis, Grunberg and others. In
particular the non-perturbative component has been used as a
regulator.
\item
As stressed in particular by Gl\"uck, Reya and Vogt 
(1991) \cite{Gluck:1991ee}, the Mellin n-moment technique for the study of the 
$Q^2$-evolution of $F_2^\gamma(x,Q^2)$ and of quark and gluon distributions 
in the photon  is
technically superior to the Bjorken-x space technique
developed earlier by Gl\"uck and Reya, Rossi, Drees, Da Luz
Vieira, Storrow and others. In particular $\alpha_s$-counting
problems can be straightforwardly avoided.
\item  
A large number of parametrizations of parton
distributions including heavy flavours have been proposed.
Compilations of these parametrizations (more than 25 in
total) can be found in \cite{Nisius,Krawczyk:2000mf}. Here the group of Marysia
Krawczyk is among the leading groups. The most recent
efforts in these directions can be found in 
\cite{Cornet:2004nb,Abramowicz:2005yd,Slominski:2005bw,Aurenche:2005da}, 
where numerous references to earlier literature can be found.
\item
The NNLO corrections have been recently completed \cite{VOGT}. I am
told that they are small but we have to wait until the
numerical analysis of these corrections has been published.
\end{itemize}

While definitely a significant progress on the study of the
implications of the Witten's and our calculations has been
done since 1978, the main progress in this field has been
done by experimentalists. After the first measurements of $F_2^\gamma(x,Q^2)$
by the PLUTO collaboration in 1981~\cite{PLUTO81} there was a dramatic
progress in collecting data made by CELLO, JADE, PLUTO, TASSO,
TOPAZ, AMY, DELPHI, L3, ALEPH, OPAL and TPC/2$\gamma$. The relevant
references can be found in 
\cite{Cornet:2004nb,Abramowicz:2005yd,Slominski:2005bw,Aurenche:2005da}. 
As a result of these efforts
$F_2^\gamma(x,Q^2)$ and the quark distributions 
$q^\gamma(x,Q^2)$ are quite well known at present, 
while the gluon distribution $G^\gamma(x,Q^2)$ is still
poorly known. The ranges in $x$ and $Q^2$ explored so far are very
impressive $0.001\le x\le 0.9$ and $1.9 \Gev^2\le Q^2\le 780 \Gev^2$.

\begin{figure}[hbt]
\vspace{0.10in}
\centerline{
\epsfysize=3.0in
\epsffile{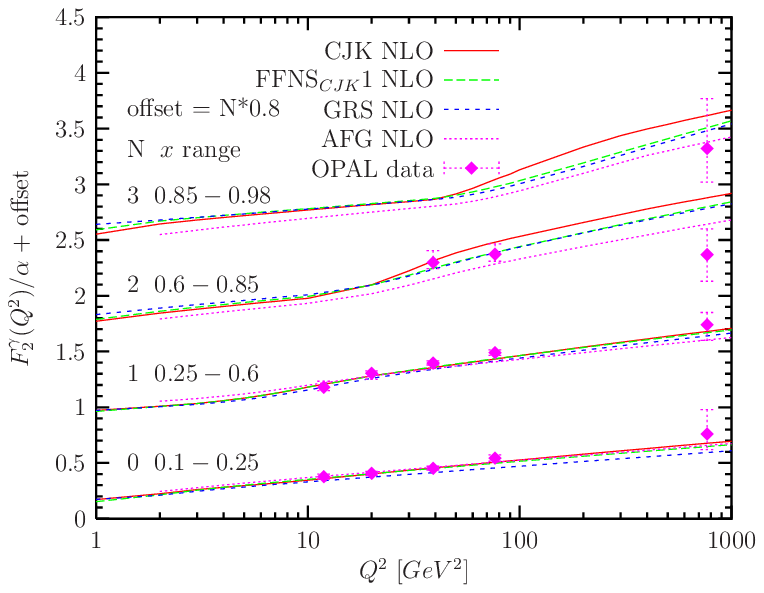}
}
\vspace{0.08in}
\caption{$F_2^\gamma$ as a function of $Q^2$ for different $x$ compared 
with experiment \cite{Cornet:2004nb}}\label{FIG3}
\end{figure}

 \section{Conclusions}
Let us finally ask whether the $\log Q^2$ and $x$-dependences of the
point-like component predicted in 1970's have been confirmed
by the data. The answer is given in Fig.~\ref{FIG3}.  Indeed the
increase of $F_2^\gamma(x,Q^2)$  with increasing $x$ and $Q^2$ 
 is clearly visible.
A more detailed analysis also shows that for $x\to 1$ the turn
over predicted by QCD becomes visible but more data is
required to expose this feature clearly.

On the other hand the present data on $F_2^\gamma(x,Q^2)$ allow already now a
rather precise determination of $\alpha_s^{\overline{MS}}$. One
finds \cite{Albino:2002ck}
\begin{equation}
\alpha_s^{\overline{MS}}(M_Z)= 0.1183\pm0.0050({\rm exp})\pm 
0.0028({\rm theory}) 
\end{equation}
in a very good agreement with the world average 
$\alpha_s^{\overline{MS}}(M_Z)= 0.1182\pm0.0027$
The accuracy could
be further improved at the NLC and a photon collider.

In summary, the photon structure functions predicted by QCD
agree well with the data, even if further studies are
clearly desirable. It became a mature field and definitely
there is more to come in the next 25 years. As I am now
returning back to flavour physics, weak decays and CP violation I wish all
explorers of the physics of photon structure functions good
luck!

\vskip0.3truecm

{\bf Acknowledgements}

\vskip0.3truecm

I would like to thank Marysia Krawczyk for inviting me to
this so well organized and very interesting symposium. The
six days I spent in Warsaw, in particular the piano
concertos in the Warsaw philharmony, will remain in my memory
forever.

\vfill\eject

\end{document}